\documentclass[sigconf, natbib=true]{acmart}

\AtBeginDocument{%
  }

\copyrightyear{2026}
\acmYear{2026}
\setcopyright{cc}
\setcctype{by}
\acmConference[WebSci Companion '26]{18th ACM Web Science Conference}{May 26--29, 2026}{Braunschweig, Germany}
\acmBooktitle{18th ACM Web Science Conference (WebSci Companion '26), May 26--29, 2026, Braunschweig, Germany}
\acmDOI{10.1145/3795513.3807433}
\acmISBN{979-8-4007-2492-3/2026/05}

\usepackage{colortbl}
\usepackage{array}
\usepackage{multirow}
\usepackage{fontawesome}
\usepackage{array}
\usepackage{subcaption}
\newcolumntype{C}[1]{>{\centering}m{#1}}

\begin{document}

\title[The WebKurator.de Platform: Combined Regional and Topical Web Curation]{The WebKurator.de Platform:\\Combined Regional and Topical Web Curation}

\author{Michael Dinzinger}
\email{michael.dinzinger@uni-passau.de}
\orcid{0009-0003-1747-5643}
\affiliation{%
  \institution{University of Passau}
  \streetaddress{Innstraße 33}
  \city{Passau}
  \country{Germany}
  \postcode{94032}
}

\author{Natanael Arndt}
\email{n.arndt@dnb.de}
\orcid{0000-0002-8130-8677}
\affiliation{%
  \institution{Deutsche Nationalbibliothek}
  \streetaddress{?}
  \city{Leipzig}
  \country{Germany}
  \postcode{?}
}

\author{Ben Böck}
\email{boeck09@ads.uni-passau.de}
\orcid{TODO}
\affiliation{%
  \institution{University of Passau}
  \streetaddress{?}
  \city{Passau}
  \country{Germany}
  \postcode{?}
}

\author{Jelena Mitrovi\'{c}}
\email{jelena.mitrovic@uni-passau.de}
\orcid{0000-0003-3220-8749}
\affiliation{%
  \institution{University of Passau}
  \streetaddress{Innstraße 33}
  \city{Passau}
  \country{Germany}
  \postcode{94032}
}

\author{Michael Granitzer}
\email{michael.granitzer@uni-passau.de}
\orcid{0000-0003-3566-5507}
\affiliation{%
  \institution{University of Passau, and IT:U Austria}
  \streetaddress{Innstraße 33}
  \city{Passau}
  \country{Germany}
  \postcode{94032}
}



\renewcommand{\shortauthors}{M. Dinzinger, N. Arndt, B. Böck, J. Mitrovi\'{c}, and M. Granitzer}

\begin{abstract}
The systematic curation of the Web remains a central challenge for national libraries and memory institutions that aim to preserve culturally and regionally relevant content. Existing directory-based approaches such as Curlie implement a predominantly topic-centric, one-dimensional hierarchy, where geographic aspects are intertwined with topical and linguistic categories.
To address this limitation, we present \textbf{WebKurator.de}, a collaborative platform for combined regional and topical web curation, initially focused on the German web. WebKurator introduces a two-dimensional curation model that explicitly separates topical categorization and geographic annotation. The system integrates LLM–based topic classification and imprint-based address extraction with geocoding, and supports user suggestions together with moderated review.
The platform is bootstrapped from the \textit{German Imprints Dataset}, a large-scale collection of 5.54 million websites. Among them, 3.14 million contain imprint pages, for which we successfully extracted and geocoded postal addresses. Of these, 2.58 million (85.17\%) are located in Germany and also have an assigned topic label. These websites form the initial foundation of WebKurator.de and can be continuously extended through user suggestions.%
\end{abstract}

\ccsdesc[500]{Information systems~Web searching and information discovery}
\ccsdesc[500]{Information systems~Web mining}

\keywords{Web Curation, Selective Web Archiving, Geographic Information Extraction, Web Directories}


\maketitle

\section{Introduction}

The Web has become a primary source for documenting cultural, economic, and social activity, motivating national libraries and memory institutions to systematically collect and preserve web content. In Germany, the Deutsche Nationalbibliothek (DNB) strives to maintain web archive collection under a comprehensive legal collection mandate. This mandate roughly covers, among others, webpages published in Germany, in German language, and content that relates to Germany.
A central challenge in this context is the curation process of relevant websites, particularly when both topical and regional coverage are required.

Historically, web directories such as the Open Directory Project (ODP/DMOZ) and its successor Curlie addressed this problem through community effort.
These projects demonstrated the value of curated web collections for academic research and beyond. However, they also revealed inherent limitations: manual curation is labor-intensive and typically one-dimensional, focusing primarily on topical hierarchies. While Curlie provides a Regional top-level category and allows websites to be listed in multiple categories, the resulting hierarchical structure remains unintuitive for queries that combine topic and geography, e.g., ``delicatessen shops in the city of Brunswick, Lower Saxony.'' As a result, the directory exhibits systematic incompleteness with respect to combined topical–regional classification, as websites are often curated along only one of the two dimensions.
Such an example is illustrated in the following:

\begin{figure}[h]
\centering
\sffamily
\vspace{-0.2cm}
\begin{tabular}{c}
\colorbox{orange!50}{http://www.bachl-feinkost.de/} \\
\begin{tabular}[c]{@{}l@{}}\: \includegraphics[width=0.28cm]{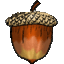} Regional > Europe > Germany > Bavaria > Cities and \\\: \hspace{0.3cm} Communities > P > Passau > Business\end{tabular} \\ [10pt]
\colorbox{orange!50}{http://www.popp-feinkost.de/} \\
\: \includegraphics[width=0.28cm]{images/curlie_nut.png} Business > Food and Related Products > Delicatessen \\
\end{tabular}
\vspace{-0.3cm}
\end{figure}

At the same time, recent advances in large language models (LLMs) have made it technically feasible to automate substantial parts of the curation process, including website categorization and address extraction.
However, fully automated approaches remain error-prone and insufficient for institutional-quality collections.

In this paper we present the \href{https://webkurator.de}{\textcolor{blue}{\emph{WebKurator}}} platform and the \emph{German Imprints Dataset}.\footnote{\url{https://openwebindex.eu/corpora/4057d6a0-0bd9-11f1-89ba-02a47ca5d9fd}}
\emph{WebKurator} is a regional and topical web curation platform motivated by the collection mandate of the DNB and serving as a prototype to revamp its selective web archiving infrastructure.
It refines the directory-style curation paradigm and process of Curlie in three important ways.
First, it introduces a two-dimensional curation model separating topic and region;
second, automation of the topic classification and geographic annotation;
third, supporting user suggestions and content moderation.
Automation is primarily realized through LLM-based topic classification and address extraction from imprint pages, followed by geocoding.
The separation of topic and region enables precise geographic queries over curated websites while maintaining a structured topical hierarchy.
Further, human moderators can monitor, validate and correct the submitted user suggestions together with their automatically generated metadata.
The architecture of WebKurator is designed to be integrated with the web archiving pipelines of national libraries.
%
The \emph{German Imprints Dataset} was created in an initial collection phase to bootstrap the platform.
It contains German websites enriched with extracted addresses, topical labels, and geo-coordinates, which are exposed and interactively curated within WebKurator.%
\footnote{The code for the collection of the German Imprints Dataset is provided on GitHub: \url{https://github.com/padas-lab-de/imprints-collection}}


In the remainder of this paper, we describe the platform design, the initial data collection, and the automated annotation workflow.



\section{Related Work}

\subsection{Human-Curated Web Directories}

Early web discovery heavily relied on human-curated web directories, most prominently the Open Directory Project (ODP), commonly referred to as DMOZ.\footnote{\url{https://dmoz.co.uk}} ODP/DMOZ organized websites into a large hierarchical taxonomy maintained by volunteer editors and served as a foundational resource for search engines and web research. Despite its success, the project struggled with scalability and editorial overhead and was discontinued in 2017.
Curlie\footnote{\url{https://curlie.org}} continues the directory-based curation approach as a community-driven successor to the project, reusing and extending its data under an open license. It preserves the core principles of manual review, editorial control and hierarchical topic classification. Despite its diminishing relevancy in the context of web discovery, Curlie remains relevant in academic contexts as a curated reference dataset, for example in website classification and categorization tasks~\cite{Lugeon2022, Hendriksen2024}.

Curlie’s underlying data model is largely topic-centric, with regional information represented implicitly through the directory hierarchy.
Concretely, Curlie distinguishes between English-language topical categories (e.g., Arts, Business, Society), which appear as top-level branches, and geographically scoped content, which is organized under separate Regional branches (e.g., North America, Europe). Websites in languages other than English are grouped under the /World category, with language-specific subtrees such as /World/Deutsch or /World/Français. This structure is the result of historical design decisions in DMOZ and persists in Curlie for backward compatibility, even though the user interface attempts to hide these technical distinctions.
As a consequence, topic, region and content language are not modeled as independent, orthogonal facets, but are instead intertwined in a single hierarchical tree.



\subsection{Selective Web Archiving Platforms}

Beyond directories, selective web archiving systems provide curated access to preserved web content rather than the live web. The Internet Archive offers large-scale crawling and access through the Wayback Machine,\footnote{\url{https://archive.org/}} while Archive-It enables institutions to curate focused web collections by selecting seeds, defining crawl scopes, and publishing collection landing pages.\footnote{\url{https://www.archive-it.eu/}}

Open-source platforms such as the Web Curator Tool (WCT)~\footnote{\url{https://webcuratortool.org/}} and NetarchiveSuite~\footnote{\url{https://github.com/netarchivesuite/netarchivesuite}} support institutional web archiving workflows, including selection, scheduling, harvesting, quality assurance, and storage in WARC format. These systems emphasize collection management and capture workflows, often assuming that seed selection and metadata creation are performed externally or manually.











\section{Collaborative Platform}

WebKurator builds upon the directory-based approach established by DMOZ and Curlie but explicitly addresses their limitations by (i) separating topical and regional dimensions, (ii) integrating automated address extraction, geocoding, and topic classification of newly suggested website entries, and thereby (iii) supporting quality control through moderation.

\begin{figure}[t]
  \centering
  \includegraphics[width=\linewidth]{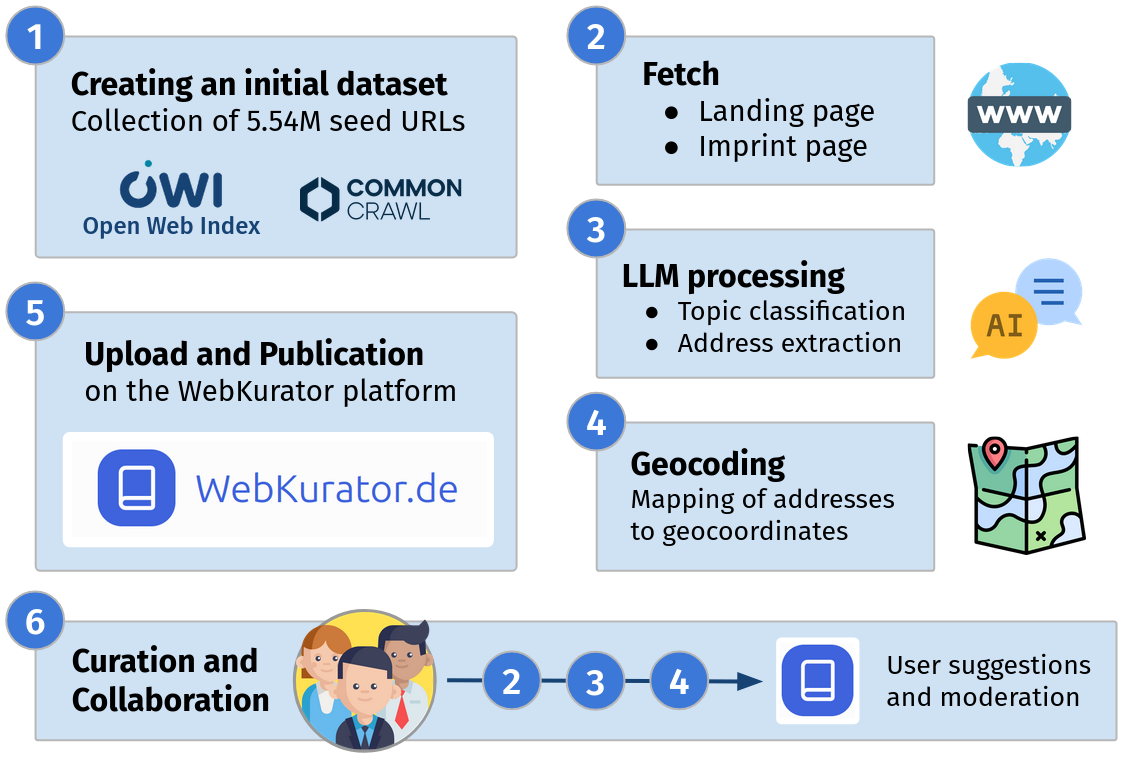}
  \caption{Overview of the WebKurator data collection and annotation workflow. Steps (1)--(5) describe the creation of the initial dataset, while step (6) shows the integration of newly suggested websites through user-driven curation.}
  \label{fig:data_collection}
  \Description{}
\end{figure}

\begin{table*}[t]
  \centering
  \caption{Comparison of top-level categories (topic) between Curlie and the German Imprints Dataset used for WebKurator.}
  \label{tab:curlie_topic_comparison}
  \vspace{-0.1cm}
  \begin{tabular}{lr lr lr lr lr}
    \toprule
    \multicolumn{6}{c}{\textbf{Curlie (English)}} &
    \multicolumn{4}{c}{\textbf{German Imprints Dataset}} \\
    \cmidrule(lr){1-6} \cmidrule(lr){7-10}
    Category & \% & Category & \% & Category & \% &
    Category & \% & Category & \% \\
    \midrule
    Business & 16.54
    & Computers & 7.33
    & Kids \& Teens & 3.41
    & Business and Prof. Services & 41.60
    & Travel and Transportation & 5.15 \\
    Arts & 16.26
    & Recreation & 6.73
    & Games & 2.71
    & Community \& Government & 16.24
    & Sports and Recreation & 4.21 \\
    Society & 15.81
    & Shopping & 5.48
    & Home & 1.69
    & Arts and Entertainment & 12.17
    & Dining and Drinking & 2.45 \\
    Science & 7.45
    & Reference & 4.40
    & News & 0.65
    & Retail & 10.65
    & Landmarks and Outdoors & 1.35 \\
    Sports & 7.38
    & Health & 4.13
    &  &
    & Health and Medicine & 5.27
    &  &  \\
    \bottomrule
  \end{tabular}
\end{table*}

\subsection{Data Collection and Annotation} \label{data_collection}

The initial set of websites uploaded to the WebKurator platform is bootstraped from the \textit{German Imprints Dataset}. We have created this dataset in an initial large-scale collection phase and it may be continuously extended through user suggestions. The initial dataset was gathered using the Open Web Crawler (OWLer) software tool~\cite{Dinzinger2024}. Seed URLs were obtained from three major web data sources, the Open Web Index~\cite{Hendriksen2024}, Common Crawl,\footnote{\url{https://commoncrawl.org/}} and Curlie.
From these sources, we selected websites that contain an imprint page. Such pages were identified by characteristic URL paths such as \texttt{/imprint} or \texttt{/impressum}.

For each discovered seed host (e.g., \texttt{www.braunschweig.de}), we downloaded the landing page and the corresponding imprint page. Since imprint pages are legally required for most German websites, they provide reliable access to structured information about the responsible organization, including postal addresses. If available, we also retrieved related pages such as contact, about-us, privacy policy, and terms of service pages for possible future use cases. We identified page types based on URL patterns (e.g., \texttt{/imprint} or \texttt{/impressum}) and anchor text. Figure~\ref{fig:data_collection} illustrates the complete workflow, including collection of seed URLs, downloading, processing and publication on the platform. Steps (1) to (5) represent the creation of the initial dataset.

After downloading the pages, we applied an automated annotation pipeline to extract structured metadata in steps (2)-(4).
Postal addresses are extracted using the locally deployed large language model \texttt{Qwen3-Next-80B}~\cite{Gadziomski2025, Qwen3-2025}. We use a \href{https://github.com/padas-lab-de/imprints-collection/blob/master/prompts/address_extraction/german/system.txt}{simple German prompt (\faLink)} that instructs the model to identify and return postal addresses from imprint pages. In addition, the pipeline performs topical classification based on a curated subset of the FourSquare taxonomy, again using \texttt{Qwen3-Next}. The FourSquare\footnote{\url{https://docs.foursquare.com/data-products/docs/categories}} hierarchy provides fine-grained, real-world–oriented categories well suited for local web content. For geocoding, we operate a local instance of Photon, an open-source geocoding service based on OpenStreetMap and Open\-Search.\footnote{\url{https://photon.komoot.io/}, \ \url{https://openstreetmap.org/} and \ \url{https://opensearch.org/}} This setup allows us to process addresses locally and provides independence from external services.

The same processing steps are applied when new websites are suggested by users. This continuous extension through user-driven, collaborative curation is illustrated in step (6) of Figure~\ref{fig:data_collection}. In this case, the system automatically downloads the relevant pages, extracts addresses, assigns topical categories, and prepares metadata before presenting the result to a moderator. Moderators can then review and correct the title, description, category assignment, extracted address, and extracted geopositions. This ensures that automated extraction supports, but does not replace, human judgment.


\begin{table}[h]
\centering
\caption{Statistics overview of the German Imprints Dataset.}
\label{tab:imprints_dataset_overview}
\vspace{-0.1cm}
\begin{tabular}{lrr}
\toprule
\textbf{Metric} & \textbf{Count} & \textbf{Perc.} \\
\midrule
Total number of downloaded websites & 5\,543\,968 & 100\% \\ \midrule
\multicolumn{3}{l}{Structured data annotations:} \\
\, $\hookrightarrow$ \texttt{Address} & 438\,913 & 7.92\% \\
\qquad $\hookrightarrow$ \texttt{GeoCoordinates} & 97\,721 & 1.76\% \\ \midrule
\multicolumn{3}{l}{Top TLDs: \texttt{.de} (52.31\%), \texttt{.com} (18.15\%), \texttt{.at} (3.70\%), \texttt{.ch} (3.58\%)} \\  
\midrule
Websites including an \textbf{imprint} & 3\,226\,331 & 58.20\% \\
\, $\hookrightarrow$ extracted address & 3\,225\,347 & 58.18\% \\
\qquad $\hookrightarrow$ assigned geoposition & 3\,142\,142 & 56.68\% \\ \midrule
\multicolumn{3}{l}{Top Countries: GER (85.57\%), AUT (7.25\%), CH (5.19\%)} \\  
\bottomrule
\end{tabular}
\end{table}

\subsection{Statistics}
The \textit{German Imprints Dataset} comprises 5.54 million websites in total, that all deliver a valid landing/home page. Of those, 3.23 million websites have an imprint.
The ccTLD \texttt{.de} is the most prevalent top-level-domain with 52.31\%. As outlined above, our approach relies on regional categorization based on address extraction from imprint pages. Table~\ref{tab:imprints_dataset_overview} highlights why this step is necessary. Although some landing pages contain structured data annotations provided by the content creators, such metadata rarely includes reliable address information or geographic coordinates. In contrast, imprint pages provide a much more consistent source of geographic information. We identified more than 3 million imprint pages, and for almost all of them we were able to extract a postal address using our LLM-based approach. For nearly all extracted addresses, we successfully assigned geographic coordinates using Photon. The final set comprises 3.14 million websites with assigned geographic positions. The majority of these websites (85.57\%) are located in Germany, followed by Austria (7.25\%) and Switzerland (5.19\%).




Table~\ref{tab:curlie_topic_comparison} compares the distribution of top-level categories in Curlie with the corresponding high-level categories derived from the German Imprints Dataset. While Curlie shows a relatively balanced distribution across traditional editorial categories such as \textit{Business}, \textit{Arts}, and \textit{Society}, our dataset is more strongly dominated by economically oriented categories, in particular \textit{Business and Professional Services}. This difference reflects the imprint-based collection strategy, which naturally emphasizes legally registered organizations. 

\subsection{Platform Functionality}





WebKurator is implemented as a collaborative web platform that operationalizes the two-dimensional curation model described above. While the underlying dataset provides large-scale coverage of the German web, the platform layer enables users to explore and refine the existing entries, and suggest new websites.

\begin{figure}[t]
  \centering
  \includegraphics[width=\linewidth]{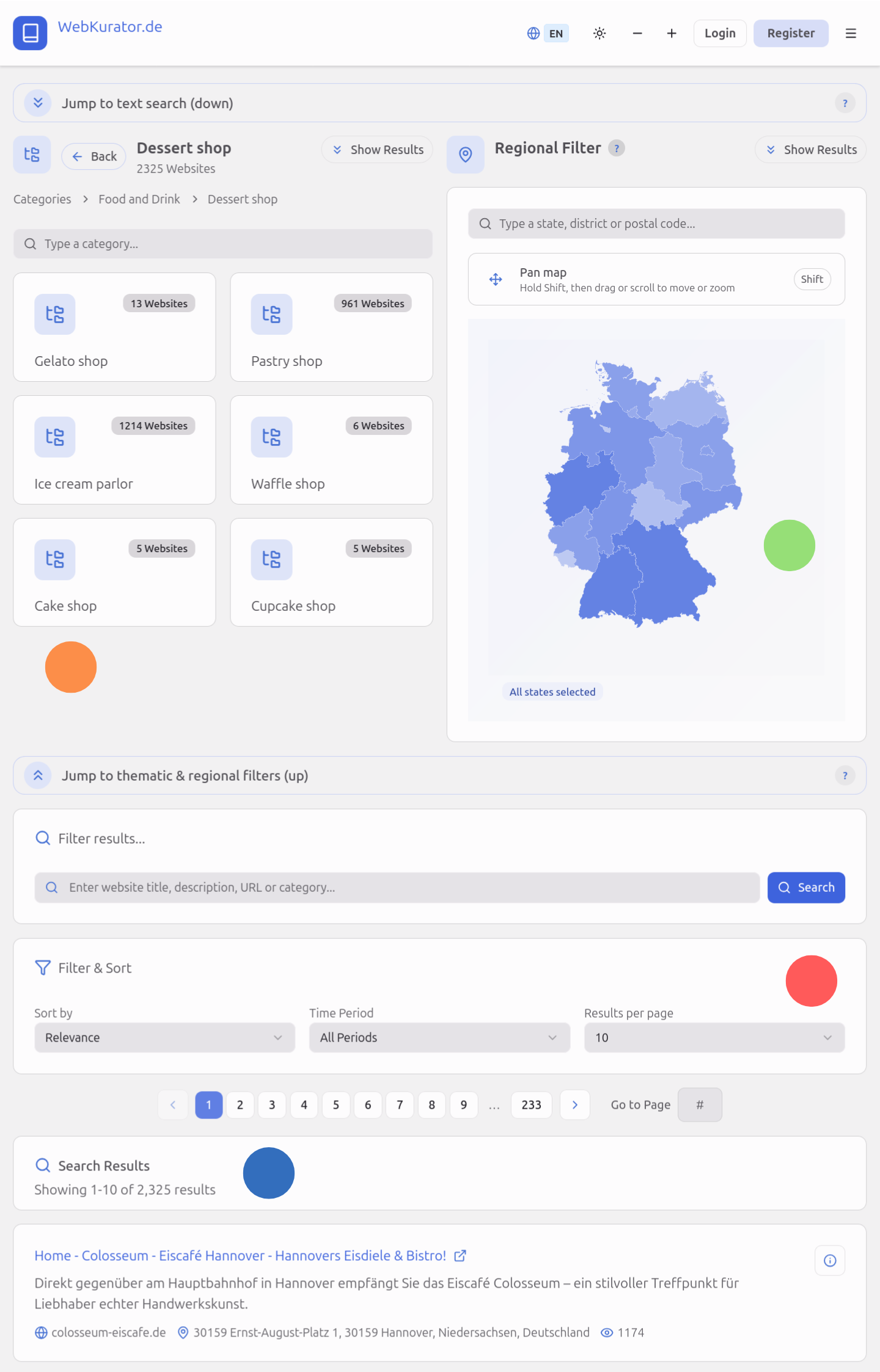}
  \caption{Search interface of WebKurator.de. The upper section provides separate filters for \textcolor{orange}{topical navigation} (left: Category > Food and Drink > Dessert Shop) and \textcolor{green}{regional selection} (right: here set to Germany). In addition, results can be refined using a \textcolor{red}{free-text query}. The lower section displays the corresponding \textcolor{blue}{search results}. Colored points mark sections.}
  \label{fig:screenshot}
  \Description{}
\end{figure}

\paragraph{Search and Exploration.}
The platform supports full-text search over website titles, descriptions, and associated metadata. Fuzzy matching allows users to retrieve relevant entries despite minor spelling variations or incomplete queries. Search can be combined with structured filters along two orthogonal dimensions: topic and region (see screenshot in Figure~\ref{fig:screenshot}).
The GUI, including the topic taxonomy, is provided in both German and English.
Regional filtering is enabled through the geocoded imprint data and supports multiple levels of administrative granularity, including federal state (\textit{Bundesland}), administrative district (\textit{Regierungsbezirk}), county (\textit{Kreis}), and postal code. Users can therefore formulate combined queries such as \textit{“Delicatessen shops in Lower Saxony”} or \textit{“IT service providers in postal code 38100”}. This faceted search design directly addresses the structural limitations of one-dimensional directory hierarchies.


\paragraph{Website Detail View and Similarity.}
Each curated host is represented by a dedicated detail page displaying its title, description, assigned categories, extracted address, and normalized geographic coordinates. To support exploratory browsing, the platform presents a list of similar websites based on shared topical categories and geographic proximity. This feature encourages discovery within regional web ecosystems and structurally related organizations.


\paragraph{Collaborative Curation Workflow.}
WebKurator enables registered users to suggest new websites, propose edits, and contribute metadata improvements. When a new website is submitted, the system automatically triggers the annotation pipeline described in Section~\ref{data_collection}. The user interface supports address auto-completion and presents LLM-based category suggestions, which can be accepted, refined, or overridden before submission.
All user contributions enter a moderation queue. Moderators can review, edit, approve, or reject suggestions. In addition, selected user actions can be reverted by administrators, allowing the restoration of deleted or modified entries and thereby supporting transparent editorial control.


\paragraph{Statistics and Transparency.}
The platform provides aggregated statistics about the dataset and platform usage, including counts of curated websites, regional distributions, and category frequencies. Registered users have access to personal contribution statistics and profile management features. Global broadcast messages and system notifications facilitate communication between administrators and the community.


\paragraph{Administration and Extensibility.}
Administrative functionality includes user management, category management, broadcast configuration, and system-level settings. The current deployment focuses on the German subset of the imprints dataset. However, the underlying architecture is designed to support expansion to additional German-speaking countries such as Austria and Switzerland, for which substantial imprint-based coverage is already available.



\section{Conclusion}

In this work, we presented WebKurator.de, a collaborative platform for combined regional and topical web curation. The platform addresses the limitations of traditional directory systems by explicitly separating topic and geographic information and enabling combined search along both dimensions.

WebKurator is based on an imprint-driven approach to geographic annotation and integrates LLM-based topic classification with automated address extraction and geocoding. The current version includes 2.58 million websites located in Germany with both topic labels and geographic information, forming a large foundation for regional exploration and selective web archiving.
The system combines automated metadata generation with structured human moderation. In this way, scalability through LLM support is balanced with editorial control and quality assurance.

While WebKurator currently focuses on the live German web, its architecture is designed to support future integration with selective web archiving workflows. We see the platform as a practical contribution to regional web preservation and as a basis for further research on geographically structured web ecosystems.

\begin{acks}
\begin{minipage}{0.30\linewidth}
\includegraphics[width=0.95\textwidth]{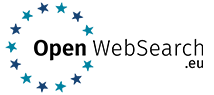}
\end{minipage}
\hfill
\begin{minipage}{0.65\linewidth}
\href{https://doi.org/10.3030/101070014}{This work has received funding from the European Union's Horizon Europe research and innovation program under grant agreement No. 101070014 (OpenWebSearch.EU)}.
\end{minipage}\\[2pt]
Furthermore, the project is supported by funds of the Federal Ministry of Agriculture, Food and Regional Identity (BMLEH) based on a decision of the Parliament of the Federal Republic of Germany via the Federal Office for Agriculture and Food (BLE) under the strategy for digitalisation in agriculture.
\end{acks}



\bibliographystyle{ACM-Reference-Format}
\bibliography{sources}



\end{document}